\documentclass [12pt]{article}
\pdfoutput=1

\usepackage{amsmath}
\usepackage{amsfonts}
\usepackage{amscd}
\usepackage{amsthm}
\usepackage{setspace}
\usepackage{amssymb}

\usepackage{graphicx}
\usepackage{authblk}
\usepackage{caption}
\usepackage{ytableau}
\usepackage{mathtools}

\def\Z{\mathbb{Z}}

\def\C{\mathbb{C}}
\def\P{\mathbb{P}}

\begin{document}

\begin{titlepage}

\begin{flushright}
KEK-TH-2267
\end{flushright}

\vskip 3.0cm

\begin{center}

{\large Four-dimensional $N=1$ theories, S-fold constraints on T-branes, \\ and behaviors in IR and UV}

\vskip 1.2cm

Yusuke Kimura$^1$ 
\vskip 0.6cm
{\it $^1$KEK Theory Center, Institute of Particle and Nuclear Studies, KEK, \\ 1-1 Oho, Tsukuba, Ibaraki 305-0801, Japan}
\vskip 0.4cm
E-mail: kimurayu@post.kek.jp

\vskip 2cm
\abstract{We analyze four-dimensional (4d) $N=1$ superconformal field theories (SCFTs) obtained as deformations of 4d $N=2$ SCFTs on S-folds by tilting 7-branes. Geometric compatibility with the structures of S-folds constrains the forms of T-branes. As a result, brane monodromies are constrained. We also discuss two 4d $N=1$ theories on probe D3-branes, where the two theories behave identically in IR, but they originate from different theories in UV. Studying the global structure of their geometry is useful in constructing these two theories.}  

\end{center}
\end{titlepage}

\tableofcontents
\section{Introduction}
\par There have been numerous advances in four-dimensional (4d) $N=2$ supersymmetric gauge theories. Seiberg and Witten obtained the exact prepotential for low-energy effective theories \cite{SeibergWitten199407, SeibergWitten199408}, which is among the major advances in this field. There have been vigorous efforts in the field since this progress; for example, \cite{Argyres1994, Argyres1995, Argyres199511, Eguchi1996, Minahan199608, Minahan199610, Argyres2007, Gaiotto2009, Cecotti2010, Argyres2015, Xie2015, Argyres201611, Caorsi2018, Borsten2018, Apruzzi2020, Argyres202003, He202004, Bourget2020, Giacomelli202007, Heckman2020, Giacomelli202010}.
\par 4d $N=2$ superconformal field theories (SCFTs) have recently been studied by utilizing S-fold constructions \cite{Borsten2018, Apruzzi2020}. A few years before these studies, the authors of \cite{Garcia-Etxebarria2015} constructed 4d theories with $N=3$ supersymmetry \footnote{Properties of 4d $N=3$ theories were discussed in \cite{Aharony2015}. Discussions of 4d $N=3$ theories can also be found in \cite{Nishinaka2016, Aharony2016, Imamura2016, Imamura201606, Agarwal2016, Lemos2016, vanMuiden2017, Bourton2018, Borsten2018, Arai2018}.} by considering finite cyclic groups $\Z_k$ acting on torus fibrations, wherein the $\Z_k$-actions also act on torus fibers. The 4d $N=2$ SCFTs studied in \cite{Apruzzi2020} flow to 4d $N=3$ theories, as constructed in \cite{Garcia-Etxebarria2015}, by mass deformations. 
\par The subjects of study in this note are 4d $N=1$ SCFTs on S-folds \footnote{See also \cite{Borsten2018} for a discussion of $N=1$ S-fold.} obtained as in the context of deforming the 4d $N=3$ theories constructed in \cite{Garcia-Etxebarria2015}. It would be natural to consider deforming the 4d $N=2$ SCFTs in \cite{Apruzzi2020} to yield such superconformal field theories, and we take this approach to study 4d $N=1$ SCFTs. 
\par There are several advantages to analyzing 4d $N=1$ SCFTs in this manner. For example, the resulting 4d $N=1$ SCFTs relate to 4d $N=2$ and $N=3$ theories by construction, and they are also relevant to studies of the structures of S-folds. Furthermore, as we discuss later in this note, they are also related to T-branes \cite{CCHV2010} \footnote{Nilpotent Higgs fields on branes were discussed in \cite{DKS2003}.}. Recent progress of T-branes can be found, e.g., in \cite{DW2011, MTmatter, MT1204, AHK2013, CR2014, CR2014space, CQV2015, HHRRZ1907}.
\par We utilize the operation of tilting 7-branes in 4d $N=2$ SCFTs to yield 4d $N=1$ SCFTs, as described in \cite{Heckman201006, Heckman201009}. This operation corresponds to allowing the Higgs field $\phi$ on the 7-branes stack to have a non-zero position-dependent vacuum expectation value (vev) over the base space, and as a result a correction term is added to the superpotential \cite{Heckman201006}. Deformation generated by this operation provides an opportunity to consider T-branes when $\phi$ and $\phi^\dagger$ do not commute \cite{CCHV2010}. 
\par In the 4d F-theory background with a gauge field flux, $\phi$ and $\phi^\dagger$ do not need to commute \cite{Heckman201006}. When they do not commute, the $N=2$ supersymmetry is broken to $N=1$ \cite{Heckman201006, Heckman201009}. We study the 4d $N=1$ SCFT on the D3-brane probing a stack of 7-branes. A D3-brane probes \footnote{Studies of D3-branes probing F-theory singularities can be found, for example, in \cite{Banks1996, Douglas1996, Fayyazuddin1998, Aharony199806, Gukov199808}.} the infrared (IR) 4d $N=1$ SCFT, which corresponds to seeing a local patch of a global geometry.

\vspace{5mm}

\par There are two themes of this note. The first is to analyze 4d $N=1$ theories \footnote{For recent studies on 4d $N=1$ theories, see, for example, \cite{Bah2011, Gadde2013, McGrane2014, Maruyoshi2016, Borsten2018, Bourton2020}.} obtained as deformations of 4d $N=3$ theories, as aforementioned. In the situation where $\phi$ and $\phi^\dagger$ do not commute, there is room to consider the effect of T-branes. Because the geometry considered here is an S-fold built as the quotient of elliptic fibration with finite cyclic group $\Z_k$ actions, the equations describing T-branes are constrained to be compatible with the $\Z_k$ actions. 
\par A T-brane is given by a spectral equation \cite{Hayashi2009, Donagi200904} that yields a multifold cover of a stack of 7-branes, where the 7-brane stack is given by $z=0$. Because the parameter $z$ transverse to the stack of 7-branes now has an equivalence relation
\begin{equation}
\label{equiv rel in intro}
z \sim e^{\frac{2\pi i}{k}}\, z
\end{equation}
under the $\Z_k$ action, a consistency condition is imposed on the spectral equation of T-branes to make them compatible with the equivalence relation (\ref{equiv rel in intro}). A physical consequence of this compatibility argument is that brane monodromies \cite{Hayashi2009, Bouchard2009, Heckman200906, Marsano2009} are constrained owing to the quotient action of the $\Z_k$ groups.  
\par The other theme is to discuss the physical phenomenon where two 4d $N=1$ SCFTs in the IR limit are indistinguishable but behave quite differently in the ultraviolet (UV) limit. We utilize a global perspective of geometry in F-theory to observe this phenomenon. This phenomenon occurs because the two globally distinct geometries have locally identical structures in a local patch that a D3-brane probes, making them appear the same in IR, but their difference becomes evident in UV. 
\par The S-fold constructions discussed in \cite{Garcia-Etxebarria2015, Apruzzi2020} were obtained as $\Z_k$-actions on elliptic fibrations over an open base wherein the complex structures of the elliptic fibers are fixed over the entire base, so the complex structures of the total spaces are compatible with the $\Z_k$ actions. However, elliptic fibrations can be constructed over a closed base space wherein the torus fibers have constant fixed complex structures over the base. Some nontrivial examples of such global geometries can be found, such as those in \cite{Dasgupta1996, Kimura2015, Kimura201603, Kimura201607, Kimura201810, Kimura201902}. Here, global geometries represent elliptic fibrations over compact base spaces \footnote{For example, an elliptic fibration over $\C^3$ is a ``local'' geometry because $\C^3$ is not closed; in contrast, as a base, an elliptic fibration over $\P^3$  describes a global geometry because $\P^3$ is closed.}. Such global geometries have some applications in 4d $N=1$ theories. We can observe a physical phenomenon that might be interesting: one can construct two global geometries on which physics appear completely identical in the IR limit, but in the UV limit, they describe different physical theories. This is owing to the fact that while the two geometries have different structures globally, their geometric structures appear locally identical. Physically, in the IR limit, they yield two identical 4d $N=1$ SCFTs probed by D3-branes that flow to very different theories in UV.  
\par In recent years, there has been progress in the appearance of discrete gauge groups in F-theory, e.g., in \cite{BM, MTsection, AGGK, KMOPR, GGK, MPTW, MPTW2, BGKintfiber, CDKPP, LMTW, Kimura2015, Kimura201603, Kimura201607, Kimura201608, Kimura201905, Kimura201907, Kimura201908}. The compactification spaces in F-theory, wherein a discrete gauge group arises, are genus-one fibrations that ``lack a global section'' \cite{MTsection}. By tuning the coefficients as parameters of the moduli of such geometries without a global section, one can deform the geometry to an elliptic fibration with global sections. In the moduli space, physics on a genus-one fibration lacking a global section and physics on an elliptic fibration with global sections are very different; a discrete gauge symmetry arises in the former geometry, while a discrete gauge symmetry does not form and U(1) does form in the latter geometry. If we consider inserting a D3-brane probe into this geometric deformation, the characteristic physical phenomenon that we mentioned previously can be observed: the D3-brane is insensitive to the subtle deformation in geometry from genus-one fibration without a global section into an elliptic fibration with sections. When one moves to UV, the effect of the global structure of the geometry becomes explicit, and the corresponding physical theories before and after the deformation behave very differently. Our argument in section \ref{sec3} suggested that a U(1) factor coming from the global sections of the elliptic fibration and a discrete gauge group arising from the geometry of the genus-one fibration lacking a global section are not reflected in the flavor symmetry group of the 4d theory on the probe D3-brane.
\par As a byproduct of the discussion of the second theme, we observe a puzzle in 4d SCFT concerning a subtlety in the structure of the flavor symmetry group. An elliptic fibration has the notion of a ``global section.'' This can be seen as a copy of the base space embedded in the total elliptic fibration \cite{MV2}. If an elliptic fibration admits a global section, global sections form an Abelian group known as a ``Mordell--Weil group.'' This group decomposes into a direct sum of free part $\Z^r$ and torsion part $\prod_i \Z_i$. In F-theory compactification \cite{Vaf, MV1, MV2} \footnote{Local models of F-theory model constructions are discussed in \cite{DWmodel, BHV, BHV2, DWGUT}. Studies of the compactification geometry of F-theory from the global perspective can be found, e.g., in \cite{MorrisonPark, MPW, BMPWsection, AL, MM, LSW, CKPT, MPT1610, Kimura1802, MizTani2018, LW1905, Kimura1910, CKS1910, Chabrol2019, Kimura1911, FKKMT1912, Kan2020, Karozas2020, Angelantonj2020, Kuramochi2020, He202009, Clingher2020}.}, the structure of the Mordell--Weil group is reflected in the gauge group structure; the rank of the Mordell--Weil group \footnote{The rank of the Mordell--Weil group is the number of copies of $\Z$ in the free part.} yields the number of U(1)s formed in F-theory \cite{MV2}, and the global structure of the gauge group is divided by the Mordell--Weil torsion \cite{AspinwallGross, AMrational, MMTW} \footnote{Computations of the Mordell--Weil torsions of elliptic fibrations in F-theory compactifications and the global structures of the gauge groups that form on the 7-branes are also discussed in \cite{Kimura201603, Kimura201607, Kimura201608, Kimura201810, Kimura201902}.}. Is the structure of such Mordell--Weil torsion dividing the flavor symmetry group observed in ``local'' 4d SCFTs probed by D3-branes? Two opposing answers to this question seem possible, as we will demonstrate. Because the two views are opposite, one must choose one of the possible two viewpoints. We discuss this puzzle in section \ref{sec4}.
\par By gauge group in sections \ref{sec3} and \ref{sec4}, we mean the gauge group of spacetime when the theory is compactified on a Calabi--Yau 4-fold. In the D3-brane probe theory, this symmetry is perceived as (part of) the flavor symmetry of the 4d theory.
\par We would like to make a remark that, when we discuss UV theories in this study, they come from the D3-branes probing compact genus-one fibered Calabi--Yau 4-folds, and 4d gravity is not entirely decoupled. The flow from UV theory to IR theory is the decoupling limit. When we consider a limit at which the size of the Calabi-Yau space becomes large enough, however, the theory approaches the situation where the D3-brane probes a non-compact space and the gravity decouples. Considering this limit if necessary, we do not consider the effect of gravity in this note. 
\par When gravity is not decoupled, because the 7-branes are wrapped on a compact space, the gauge field on the 7-branes is dynamical. Thus, the symmetries on the 7-branes yield the gauge groups in UV theories. Since the gauge field is decoupled in the decoupling limit, in the D3-brane probe theory in the IR the symmetry on the 7-branes is perceived as the flavor symmetry of the 4d theory.

\vspace{5mm}

\par In section \ref{sec2.1}, we briefly review the 4d $N=3$ theories constructed in \cite{Garcia-Etxebarria2015} and the 4d $N=2$ SCFTs realized in F-theory on S-folds in \cite{Apruzzi2020} as deformations of 4d $N=3$ theories. We construct and analyze 4d $N=1$ SCFTs as deformations of these theories in section \ref{sec2.2}. Tilting 7-branes deforms 4d $N=2$ SCFTs, which yields 4d $N=1$ SCFTs in IR \cite{Heckman201009}. After we discuss this construction, we analyze brane monodromies in the resulting 4d $N=1$ theories in section \ref{sec2.2}. The structures of T-branes are constrained by the symmetry of S-fold constructions. We analyze the constraints that brane monodromies receive as a consequence of this.
\par We discuss in section \ref{sec3} two 4d $N=1$ theories that behave in an identical manner in IR but flow to very distinct theories in UV. In these constructions, considering the global aspects of the geometry is essential. When a similar argument is applied to the global structure of the gauge group formed on 7-branes, we observe a puzzle, which we discuss in section \ref{sec4}. We present our concluding remarks in section \ref{sec5} alongside some open problems.

\section{Brane monodromies in 4d $N=1$ theories as deformations of 4d $N=2$ SCFTs on S-folds}
\label{sec2}

\subsection{Reviews of 4d $N=3$ theory and 4d $N=2$ SCFTs on S-folds}
\label{sec2.1}
\par We first briefly review the 4d $N=3$ theories constructed in \cite{Garcia-Etxebarria2015}. We then review the 4d $N=2$ SCFTs on S-folds constructed in \cite{Apruzzi2020}. We will discuss 4d $N=1$ SCFTs obtained as deformations of these theories, and T-branes and brane monodromies in the resulting 4d $N=1$ theories in section \ref{sec2.2}. We review the 4d $N=2$ and $N=3$ theories in this section, as we require them to construct 4d $N=1$ theories. 
\par F-theory compactification on the product $\C^3\times T^2$ yields a 4d $N=4$ theory. The coordinates of $\C^3$ are denoted by $z_1, z_2$, and $z_3$, and the complex coordinate of the two-torus in the product is denoted by $z_4$. The authors of \cite{Garcia-Etxebarria2015} considered the following action of a finite cyclic group $\Z_k$ on the product $\C^3\times T^2$:
\begin{eqnarray}
\label{N=3 action in 2}
z_1 \rightarrow & e^{\frac{2\pi i}{k}}z_1 \\ \nonumber
z_2 \rightarrow & e^{-\frac{2\pi i}{k}}z_2 \\ \nonumber
z_3 \rightarrow & e^{\frac{2\pi i}{k}}z_3 \\ \nonumber
z_4 \rightarrow & e^{-\frac{2\pi i}{k}}z_4.
\end{eqnarray}
Here, the complex structure $\tau$ of the two-torus as an elliptic fiber is constant over the base $\C^3$, so the product $\C^3\times T^2$ is a direct product as a complex manifold. Because the $\Z_k$-action acts on the two-torus as a fiber, it is necessary to confirm that the action is compatible with the elliptic fibration structure of $\C^3\times T^2$. It was shown in \cite{Garcia-Etxebarria2015} that the compatibility conditions require the order $k$ of the cyclic group $\Z_k$ to take one of the values $k=2,3,4$, or $6$, and the complex structure of the elliptic curve as a fiber must take a specific value for each of the values of $k$ \cite{Garcia-Etxebarria2015}. 
\par We describe the complex structures of elliptic fibers of an elliptic fibration in detail here. An elliptic curve (two-torus seen as a complex curve) is described by a Weierstrass equation. A Weierstrass equation is an equation of the following form:
\begin{equation}
y^2=x^3+f\, x+g.
\end{equation}
Then, the complex structure of an elliptic curve as a fiber of the fibration is uniquely labeled by a j-invariant, which is an invariant under the isomorphisms of elliptic curves. A j-invariant $j$ is given in terms of the Weierstrass coefficients $f,g$, as follows \cite{Cas}:
\begin{equation}
j= \frac{1728\cdot 4\, f^3}{4f^3+27g^2}.
\end{equation}
\par Modular parameter $\tau$ also specifies the complex structure of an elliptic curve. $\tau$ takes a value in $\C$ modulo $SL(2,\Z)$-action. 
\par When the order $k$ of the group $\Z_k$ takes the values $k=3,4,6$, $N=4$ supersymmetry breaks down to $N=3$, and as a result of the orbifold action (\ref{N=3 action in 2}), a 4d $N=3$ theory is obtained \cite{Garcia-Etxebarria2015}. 

\vspace{5mm}

\par Before we review 4d $N=2$ SCFTs on S-folds, we make a remark. Construction of a genus-one fibration whose elliptic fiber has a constant complex structure throughout the base, that is {\it not} a direct product of a two-torus as a fiber and the base space, is possible. Genus-one fibered K3 surfaces, whose fibers have constant complex structures $\tau=exp(\frac{\pi i}{3})$ and $\tau=i$ over the base $\P^1$ (equivalently, they have j-invariants 0 and 1728, respectively) can be found in \cite{Kimura2015, Kimura201603} \footnote{Higher dimensional generalizations to genus-one fibered Calabi--Yau 3-folds and genus-one fibered Calabi--Yau 4-folds can be found in \cite{Kimura201607, Kimura201810, Kimura201902}.}. These global geometries play a role in the construction of 4d $N=1$ theories in section \ref{sec3}. 

\vspace{5mm}

\par The authors of \cite{Apruzzi2020} constructed 4d $N=2$ SCFTs by considering F-theory compactifications on S-folds as deformations of the 4d $N=3$ theories constructed in \cite{Garcia-Etxebarria2015}, and we briefly review this construction. 
\par The S-folds in \cite{Apruzzi2020} were built by acting finite cyclic groups $\Z_k$ on the space $K3^{\circ}\times \C^2$. Here, $K3^{\circ}$ denotes an open patch in an elliptic K3 surface such that the base of $K3^{\circ}$ is $\C$. $K3^{\circ}$ is an elliptic fibration over $\C$. The coordinate of this base $\C$ is denoted as $z_1$, and the coordinates of $\C^2$ in the product $K3^{\circ}\times \C^2$ are denoted as $z_2, z_3$. $z_1$ yields the coordinate transverse to the stacks of 7-branes, and $z_2, z_3$ yield the coordinates parallel to the stacks of 7-branes. $z_1$ is identified with a chiral superfield, $Z_1$, which parameterizes the Coulomb branch, and $z_2, z_3$ are identified with a decoupled hypermultiplet, $Z_2, Z_3$ in the 4d theory on the probe D3-brane. The $\Z_k$-action acting on $K3^{\circ}\times \C^2$ is relaxed from condition (\ref{N=3 action in 2}) to the following condition \cite{Apruzzi2020}:
\begin{eqnarray}
\label{N=2 action in 2.1}
z_1 \rightarrow & e^{\frac{2\pi i}{k}}z_1 \\ \nonumber
z_2 \rightarrow & e^{i\, \phi_2}z_2 \\ \nonumber
z_3 \rightarrow & e^{i\, \phi_3}z_3,
\end{eqnarray}
where $\phi_2$ and $\phi_3$ are required to satisfy 
\begin{equation}
\label{phase relation in 2.1}
\phi_2+\phi_3=0 \hspace{2mm} {\rm mod}\, 2\pi n.
\end{equation}
This construction yields S-folds, and F-theory on the resulting S-folds with a probe D3-brane yields 4d $N=2$ SCFTs \cite{Apruzzi2020}.
\par Analogous to 4d $N=3$ theories, compatibility conditions are imposed on the values of $k$ and the complex structures of the elliptic fibers. The complex structure of the fibers of the elliptic fibration $K3^{\circ}$ must be constant over the base $\C$, i.e., the axio-dilaton must take a constant value over $\C^3$. $\Delta_7$ is used to represent the scaling dimension of the Coulomb branch (CB) operator for the 7-brane in $K3^{\circ}\times \C^2$. When $\Z_k$-groups act on the product to yield S-folds, the authors of \cite{Apruzzi2020} argued that the theories on the resulting S-folds should have the scaling dimensions $k\Delta_7$ and that these correspond to a phase rotation of supercharges. They deduced from these that $k\Delta_7=2,3,4,6$, and consequently, they found that the S-folds that are compatible with $\Z_k$-actions are as follows \cite{Apruzzi2020}: $\Z_2$ S-folds with type $H_2, D_4, E_6$ 7-branes \footnote{Here $H_l$, $l=1,2$, represents the Argyres--Douglas theories \cite{Argyres1995, Argyres199511, Eguchi1996} that arise from an $SU(2)$ theory possessing $l+1$ flavors.}; $\Z_3$ S-folds with type $H_1, D_4$ 7-branes; and $\Z_4$ S-folds with type $H_2$ 7-brane. The axio-dilaton has a constant value over the base space, and the value depends on the value of $k\Delta_7$ \cite{Apruzzi2020}.  
\par We remark here that the types of 7-branes for S-folds mentioned previously are the types before the quotient actions of $\Z_k$. The authors of \cite{Apruzzi2020} computed the $\Z_k$-quotients of the corresponding singular fiber types by analyzing the $\Z_k$-actions on the Weierstrass equations directly. 
\par However, there is another method to compute the quotient fiber types, which we discuss here. The complex structure of a singular fiber is invariant under the action of the finite cyclic group $\Z_k$. Alternatively, the j-invariant of a singular fiber is unchanged under the quotient action of $\Z_k$. Utilizing this mathematical fact, one can deduce the singular fiber types as a result of the quotient action of $\Z_k$ groups. The types of singular fibers of elliptic surfaces, including their j-invariants, were classified by Kodaira in \cite{Kod1, Kod2}. Kodaira's classification of the types of singular fibers \footnote{\cite{Ner, Tate} discussed techniques to determine the types of the singular fibers of an elliptic fibration.} for elliptic surfaces is presented in Table \ref{table fibertype in 2.1}.

\begingroup
\renewcommand{\arraystretch}{1.1}
\begin{table}[htb]
  \begin{tabular}{|c|c|r|c|c|c|} \hline
brane type & J-invariant & Monodromy  & Order of Monodromy & \# of 7-branes & singularity \\ \hline
$I^*_0$ & regular & $-\begin{pmatrix}
1 & 0 \\
0 & 1 \\
\end{pmatrix}$ & 2 & 6 & $D_4$\\ \hline
$I_b$ & $\infty$ & $\begin{pmatrix}
1 & b \\
0 & 1 \\
\end{pmatrix}$ & infinite & $b$ & $A_{b-1}$\\
$I^*_b$ & $\infty$ & $-\begin{pmatrix}
1 & b \\
0 & 1 \\
\end{pmatrix}$ & infinite & $b+$6 & $D_{b+4}$\\ \hline
$II$ & 0 & $\begin{pmatrix}
1 & 1 \\
-1 & 0 \\
\end{pmatrix}$ & 6 & 2 & none.\\
$II^*$ & 0 & $\begin{pmatrix}
0 & -1 \\
1 & 1 \\
\end{pmatrix}$ & 6 & 10 & $E_8$\\ \hline
$III$ & 1728 & $\begin{pmatrix}
0 & 1 \\
-1 & 0 \\
\end{pmatrix}$ & 4 & 3 & $A_1$\\
$III^*$ & 1728 & $\begin{pmatrix}
0 & -1 \\
1 & 0 \\
\end{pmatrix}$ & 4 & 9 & $E_7$\\ \hline
$IV$ & 0 & $\begin{pmatrix}
0 & 1 \\
-1 & -1 \\
\end{pmatrix}$ & 3 & 4 & $A_2$\\
$IV^*$ & 0 & $\begin{pmatrix}
-1 & -1 \\
1 & 0 \\
\end{pmatrix}$ & 3 & 8 & $E_6$\\ \hline
\end{tabular}
\caption{\label{table fibertype in 2.1}Types of the singular fibers of elliptic surfaces and their properties \cite{Kod1, Kod2} and well-known brane interpretations. ``Regular'' for j-invariant of type $I_0^*$ indicates the fact that j-invariant of type $I_0^*$ fiber can take any finite value in $\C$.}
\end{table}
\endgroup

Because there are $\Z_2$ and $\Z_4$ S-folds with type $H_2$ 7-branes according to the results of \cite{Apruzzi2020}, $\Z_2$- and $\Z_4$-quotients of type $IV$ fibers can be considered. Type $IV$ fibers have j-invariant 0 according to Table \ref{table fibertype in 2.1}; therefore, $\Z_2$- and $\Z_4$-quotients of this fiber type must have j-invariant 0. The order of the $A_2$ singularity corresponding to type $IV$ fibers is 4, and the $\Z_2$-action doubles this order, so the resulting fiber type has a corresponding singularity type of order 8. Thus, we uniquely determine that the resulting fiber type is $IV^*$. This is consistent with the result given in \cite{Apruzzi2020}. A similar reasoning applied to the $\Z_4$-quotient appears to show that the resulting fiber type has order 16, and this appears to be a bad singularity. In fact, this is not the case. To see this, consider the following reparameterization of the Weierstrass coefficients:
\begin{equation}
f\rightarrow z^2\cdot f, \hspace{4.5mm} g\rightarrow z^3\cdot g.
\end{equation}
Because the discriminant is given by $\Delta \sim 4f^3+27g^2$, this reparameterization reduces the order of the singularity by six, and the actual order of singularity is 10. Therefore, the $\Z_4$-quotient of a type $IV$ fiber has a j-invariant whose corresponding singularity has order 10, and this is uniquely determined to be a type $II^*$ fiber. This is also consistent with the result in \cite{Apruzzi2020}.
\par A similar reasoning applies to the $\Z_k$-quotient of other fiber types. The results of the quotient fiber types obtained in \cite{Apruzzi2020} are perfectly consistent with the results deduced via the method described herein. 
\par S-fold quotient constrains the deformation parameters of the Weierstrass equation of an elliptic fibration \cite{Apruzzi2020, Heckman2020}, and the mass deformation parameters form the Casimir invariants of the resulting flavor symmetry. The flavor symmetries of 4d $N=2$ SCFTs on S-folds were discussed in \cite{Apruzzi2020, Giacomelli202007, Heckman2020}. We list the rank-one 4d $N=2$ SCFTs on S-folds with discrete torsion and their flavor symmetries deduced in \cite{Apruzzi2020} \footnote{Rank-one 4d $N=2$ SCFTs on S-folds without discrete torsion and their flavor symmetries can be found in \cite{Heckman2020}.} in Table \ref{table flavor symmetry in 2.1}.

\begingroup
\renewcommand{\arraystretch}{1.1}
\begin{table}[htb]
\centering
  \begin{tabular}{|c|c|} \hline
S-fold quotient of singular fiber & rank-one 4d $N=2$ SCFTs \\ \hline
$IV^*$, $\Z_2$ & $[II^*, \, C_5]$ \\ 
$I_0^*$, $\Z_2$ & $[III^*, \, C_3C_1]$ \\
$I_0^*$, $\Z_3$ & $[II^*, \, A_3\rtimes \Z_2]$ \\ 
$IV$, $\Z_2$ & $[IV^*, \, C_2U_1]$ \\
$IV$, $\Z_4$ & $[II^*, \, A_2 \rtimes \Z_2]$ \\
$III$, $\Z_3$ & $[III^*, \, A_1U_1 \rtimes \Z_2]$ \\ \hline
\end{tabular}
\caption{\label{table flavor symmetry in 2.1}Rank-one 4d $N=2$ SCFTs on S-folds with discrete torsion corresponding to quotients of Kodaira fibers \cite{Apruzzi2020} are listed. Items in the left column represent $\Z_n$ S-fold quotients of Kodaira fibers. For example, $IV^*, \Z_2$ indicates $\Z_2$ quotient of type $IV^*$ Kodaira fiber. The first entries in the square brackets in the right column represent the Kodaira fiber types corresponding to the Coulomb-branch singularities. The groups appearing as the second entries in the square brackets in the right column represent the flavor symmetries of the probe D3-branes in the resulting rank-one 4d $N=2$ SCFTs. We follow the notational convention utilized in \cite{Argyres201611}.}
\end{table}
\endgroup

\subsection{4d $N=1$ SCFTs as deformations of 4d $N=2$ SCFTs on S-folds, T-branes in 4d $N=1$ S-fold and brane monodromy}
\label{sec2.2}
\par One of the aims of this note is to analyze the 4d $N=1$ SCFTs obtained as deformations of the 4d $N=2$ SCFTs on S-folds \cite{Apruzzi2020} that we reviewed previously. We require an operation that breaks $N=2$ supersymmetry down to $N=1$; we utilize an operation of tilting a stack of 7-branes, as described in \cite{Heckman201006, Heckman201009}. This operation corresponds to the Higgs field $\phi$ on the 7-branes stack acquiring a non-zero position-dependent vev \cite{Heckman201006}. Then, tilting the stack of 7-branes adds a correction to the superpotential as $Tr_G(\phi(Z_1, Z_2)\cdot \mathcal{O})$ \cite{Heckman201006}. $\mathcal{O}$ denotes dimension-two operators in the D3-brane probe theory transforming in the adjoint representation of $G$, where $G$ represents the flavor symmetry group \cite{Heckman201006}. When $\phi$ and $\phi^\dagger$ do not commute, the $N=2$ supersymmetry is broken down to $N=1$ \cite{Heckman201006}. $N=1$ deformation by tilting the stack of 7-branes corresponds to the superpotential deformation $\delta W=Tr_G(\phi(Z_1, Z_2)\cdot \mathcal{O})$ as discussed in \cite{Heckman201006, Heckman201009}.  
\par The adjoint-valued vev that $\phi$ takes, in the situation $[\phi, \phi^\dagger]\ne 0$, yields room to consider the T-brane structure \cite{CCHV2010}. When the stack of $n$ 7-branes is located at $z_1=0$, the T-brane is generally given by the following spectral equation \cite{CCHV2010}:
\begin{equation}
\label{spectral T-brane in 2.1}
z_1^n+b_2z_1^{n-2}+b_3z_1^{n-3}+\ldots+b_n=0,
\end{equation}
where $z_1$ denotes the spectral parameter, and $b_2, \ldots, b_n$ are polynomials in the variables of $z_2$ and $z_3$. Equation (\ref{spectral T-brane in 2.1}) can be seen as the spectral equation for $\phi$. 
 
\par A T-brane is a structure associated with the coincident 7-branes that the Weierstrass equation cannot capture \cite{MT1204}. This physical degree of freedom in string theory is not fully determined by the defining equation of the geometry and still receives constraints from the geometry when an S-fold is considered. In other words, the structure of the T-brane must be compatible with the quotient action of the $\Z_k$ groups. We now explain this in more detail. 
\par The $\Z_k$ quotient of $K3^\circ \times \C^2$, $k=2,3,4$, was used to construct S-folds in \cite{Apruzzi2020}. $K3^\circ$ denotes an open patch on the K3 surface, as previously mentioned. We focus on these S-folds because the 4d $N=1$ theories analyzed here are obtained as deformations of the 4d $N=2$ SCFTs constructed in \cite{Apruzzi2020}. 
\par The base of $K3^\circ$ as an elliptic fibration is isomorphic to $\C$, and the coordinate of the base is denoted by $z_1$. The coordinates of $\C^2$ in the product $K3^\circ \times \C^2$ are denoted by $z_2, z_3$. $\Z_k$ action are given by (\ref{N=2 action in 2.1}) \cite{Apruzzi2020}, as mentioned in section \ref{sec2.1}, where phases $\phi_2, \phi_3$ satisfy the relation (\ref{phase relation in 2.1}). The product $z_2z_3$ is invariant under $\Z_k$ action (\ref{N=2 action in 2.1}). $z_1$ is identified with $e^{\frac{2\pi i}{k}}\, z_1$ under the quotient action. 
\par Note here that when S-fold construction is considered, geometric conditions are imposed on the equation (\ref{spectral T-brane in 2.1}). For a $\Z_k$ S-fold, the coordinate $z$ satisfies the equivalence relation
\begin{equation}
z_1 \sim e^{\frac{2\pi i}{k}}\, z_1.
\end{equation}
For simplicity, we focus on the situations where the coefficients $b_i$, $i=2, \ldots, n$, are polynomials that only depend on the product $z_2 z_3$ as a variable. Namely, we only consider the cases in which they are given as
\begin{equation}
\label{coeff b in 2.2}
b_i =\sum_j c_{ij}\, (z_2z_3)^j.
\end{equation}
As stated previously, the product $z_2z_3$ is invariant under $\Z_k$ action (\ref{N=2 action in 2.1}). Under this assumption, conditions are imposed only on powers of $z_1$ in the spectral equation (\ref{spectral T-brane in 2.1}) to make it compatible with the $\Z_k$ action. For simplicity of the argument, we focus on this situation and proceed.  
 
\par The T-brane is then given by the spectral equation of the following form:
\begin{equation}
F(z)=z_1^n+b_2(z_2z_3)\, z_1^{n-2}+b_3(z_2z_3)\, z_1^{n-3}+\ldots+b_n(z_2z_3)=0,
\end{equation}
where $b_i(z_2z_3)$, $i=2, \ldots, n$, are polynomials in $z_2z_3$. The form of the T-brane must be invariant under $\Z_k$ action (\ref{N=2 action in 2.1}). Because the coefficients $b_i(z_2z_3)$, $i=2, \ldots, n$ are invariant under $\Z_k$ action (\ref{N=2 action in 2.1}), this is equivalent to the requirement that $F(z)$ satisfies the following relation:
\begin{equation}
\label{constraint on T-brane in 2.2}
F(e^{\frac{2\pi i}{k}}\cdot z_1) = c\, F(z_1).
\end{equation}
Here, $c$ denotes a non-zero constant number. When this condition is satisfied, under the quotient action of $\Z_k$, the equation $F(z)=0$ is invariant; the form of the T-brane is unchanged under the quotient action. 
\par The condition (\ref{constraint on T-brane in 2.2}) clearly constrains the form of the polynomial $F(z)$ in the spectral equation. The physical consequence of this \footnote{Possibilities that the coefficients $b_i$ take forms different from (\ref{coeff b in 2.2}) are not excluded. The constraints imposed on the spectral equation by the compatibility with $\Z_k$ action (\ref{N=2 action in 2.1}), however, do not essentially change for such possibilities, and we expect that a physical consequence similar to that we discuss here applies to the possible situations wherein the coefficients $b_i$ take forms different from (\ref{coeff b in 2.2}) in the spectral equation.} is as follows. Because the Galois group of the polynomial $F(z)$ yields brane monodromy \cite{Heckman201009, CCHV2010}, the S-fold compatibility condition (\ref{constraint on T-brane in 2.2}) constrains the brane monodromy. When conditions are imposed on the coefficients of the polynomial $F(z)$, the associated Galois group generally becomes smaller than the symmetric group $S_n$. As a result, the brane monodromy group becomes smaller owing to the compatibility condition (\ref{constraint on T-brane in 2.2}). Although the T-brane is a structure associated with the coincident 7-branes that the Weierstrass equation cannot fully capture, its structure still receives restrictions from the geometry, as we have shown. Compatibility with the $\Z_k$ orbifold actions physically constrains the structures of T-branes and brane monodromies.

\vspace{5mm}

\par As an example, we consider the T-brane structure in 4d theory on an S-fold built from an open patch in K3 surface with an $A_2$ singularity corresponding to type IV fiber. 
\par For this case, 7-branes in the open patch of the K3 surface have type $H_2$, and the type-$H_2$ stack consists of four 7-branes. $\Z_2$ action acted on the open patch K3$^\circ$ times $\C^2$ yields $\Z_2$ S-fold. As we mentioned previously, the theory on probe D3-brane inserted in the resulting S-fold yields a 4d $N=2$ SCFT \footnote{This theory flows to a 4d theory with $N=4$ supersymmetry, instead of $N=3$ theory \cite{Garcia-Etxebarria2015, Apruzzi2020}.}, and considering T-brane structure breaks half of $N=2$ supersymmetry which yields 4d $N=1$ theory. The spectral equation of the T-brane in the resulting 4d $N=1$ theory is given by degree-four characteristic polynomial. The spectral equation must satisfy the relation (\ref{constraint on T-brane in 2.2}) under the action of $\Z_2$ group. 
\par A T-brane given by the spectral equation of the following form is invariant under the $\Z_2$-action:
\begin{equation}
\label{Z2 T-brane in 2.2}
F(z)=z_1^4+a\, z_1^2+b=0.
\end{equation}
$a,b$ are polynomials in $z_2z_3$ as we previously mentioned. The equation (\ref{Z2 T-brane in 2.2}) satisfies the relation (\ref{constraint on T-brane in 2.2}). A computation shows that the Galois group of the polynomial (\ref{Z2 T-brane in 2.2}) is isomorphic to $D_8$, the dihedral group of order 8. Therefore, the 7-brane monodromy group for the 4d $N=1$ theory obtained from the $\Z_2$ S-fold is $D_8$. 
\par Utilizing an open patch in K3 surface with an $A_2$ singularity corresponding to a type IV fiber, one can also construct a $\Z_4$ S-fold. For this construction, a T-brane described by the following equation is invariant under the $\Z_4$ action:
\begin{equation}
F(z)=z_1^4+a=0.
\end{equation}
The Galois group of this polynomial is isomorphic to $\Z_4$, and the 7-brane monodromy group for the 4d $N=1$ theory obtained from the constructed $\Z_4$ S-fold is $\Z_4$.

\vspace{5mm}

\par We make a remark on the flavor symmetries of the probe D3-branes in 4d $N=1$ SCFTs obtained as deformations of the 4d $N=2$ SCFTs on S-folds by T-branes. Because the effect of considering T-branes corresponds to tilting the stack of 7-branes, we expect that the resulting flavor symmetry algebra of the probe D3-brane in 4d $N=1$ SCFT obtained as a deformation of a 4d $N=2$ SCFT on an S-fold by T-branes is a subalgebra of the flavor symmetry algebra of that 4d $N=2$ SCFT. This argument suggests that the flavor symmetry algebras of the 4d $N=1$ SCFTs obtained as deformations of the 4d $N=2$ SCFTs on S-folds by T-branes are given as subalgebras of the flavor symmetry algebras of the 4d $N=2$ SCFTs deduced in \cite{Argyres2015, Apruzzi2020, Giacomelli202007, Heckman2020}.

\section{Two 4d $N=1$ theories behaving identically in IR but behaving differently in UV}
\label{sec3}
\par We discussed structures of some 4d $N=1$ SCFTs probed by D3-branes in section \ref{sec2}. The D3-brane only probes a local small patch of the geometry, and theories behaving identically in IR do not necessarily behave identically in UV. In this section, we aim to provide two 4d $N=1$ theories that behave identically in IR but differently in UV by analyzing the geometry from a global perspective. When the global structures of the geometries are distinct, as theories flow to the UV regime, they become sensitive to the difference in the global structures of the geometries. 
\par Our argument in this section applies to general geometries constructed as genus-one fibrations with multisections and their deformations into elliptic fibrations with multiple global sections. Our argument applies to 4d $N=1$ SCFTs constructed using S-folds, as discussed in section \ref{sec2}, as special cases of such geometries. As constructed in \cite{Kimura2015, Kimura201603}, there are genus-one fibrations lacking a global section whose fibers have constant j-invariants 0 and 1728 throughout the base spaces (or equivalently, having complex structures $\tau=exp(\frac{\pi i}{3})$ and $\tau=i$). One can construct ``closed'' $\Z_k$ S-folds \footnote{``Closed'' means that they have compact base spaces.} by acting $\Z_k$ groups on these genus-one fibrations. Because the D3-brane only probes a local open patch, theories on such closed S-folds in the IR limit reduce to 4d $N=1$ SCFTs on S-folds, as discussed in section \ref{sec2}.
\par We have stated the capacity to which our argument in this section applies. Before discussing the central point of our argument, we explain the physical meanings of multisection geometries and elliptic fibrations with multiple sections. It is known that a discrete gauge group arises on multisection geometry in F-theory \cite{MTsection}. A multisection is a multifold cover of the base space of a genus-one fibration. The times a multisection wraps around over the base is referred to as the ``degree'' of the multisection. When a multisection has degree $n$, it is concisely referred to as an ``$n$-section.'' A discrete $\Z_n$ gauge group forms in F-theory on an $n$-section geometry \cite{MTsection}. The special case of $n=1$ corresponds to a global section, yielding a copy of the base space. 
\par By tuning the coefficients of the defining equation for a genus-one fibration with a multisection to special values, a multisection splits into multiple global sections. The number of independent global sections that an elliptic fibration possesses minus one yields the number of U(1)s formed in F-theory on that elliptic fibration, as discussed in \cite{MV2}. Therefore, when an $n$-section splits into $n$ global sections, a U(1)$^{n-1}$ forms in F-theory. Tuning an $n$-section geometry to an elliptic fibration with $n$ sheets of global sections can be viewed as the reverse of the Higgsing process, wherein a model with U(1)$^{n-1}$ breaks down into a mode with a $\Z_n$ gauge group \cite{MTsection}. 
\par It would be natural to expect that because the D3-brane only probes a small local patch of the geometry, it is insensitive to tuning of a genus-one fibration with an $n$-section into an elliptic fibration with $n$ global sections. We demonstrate that this expectation is true. 
\par The key point to show this is that an $n$-section and $n$ separate global sections are locally indistinguishable. An $n$-section is an $n$-fold branched cover of the base space. In other words, an $n$-section is $n$ global sections combined along the branching loci. When the branching loci diminish and eventually disappear, the $n$-section splits into $n$ separate global sections. When one takes a small open patch  such that it avoids any point in the branching loci, that local patch does not distinguish the $n$-section and $n$ separate global sections; the only difference is whether $n$ copies of the base space are combined along the branching loci or not, but that local patch does not pass any branched point, so they locally have no difference.
\par Now we tune a genus-one fibration with an $n$-section to an elliptic fibration with $n$ separate global sections while maintaining the type of the 7-branes stack at which the probe D3-brane is placed. The probe D3-brane is thus insensitive to this deformation in the geometry, as we have demonstrated. 
\par In particular, this means that in the IR limit, the D3-brane does not probe the Higgsing process of U(1)$^{n-1}$ breaking down into a $\Z_n$ group occurring through the global geometric deformation in the moduli. U(1) or $\Z_n$ groups arising from the global feature of the geometry in F-theory are not reflected in the 4d $N=1$ SCFT in IR. 
\par For the two theories, realized as an F-theory on an $n$-section geometry and an F-theory on an elliptic fibration with $n$ global sections, the two theories on the D3-brane probes flow to the identical theory in IR. In UV, differences in the global geometric structures are probed and flow to distinct theories, reflecting the difference between U(1)$^{n-1}$ and $\Z_n$ groups.

\vspace{5mm}

\par We would like to demonstrate that it is possible to deform a multisection geometry to an elliptic fibration with multiple global sections while maintaining the type of stack of the 7-branes at which the D3-brane probes. We show this for the case of a bisection geometry \footnote{Bisection refers to an $n$-section with $n=2$.} \cite{BM, MTsection, Kimura201603}. A bisection geometry generally admits an expression as the double cover of a quartic polynomial, as discussed in \cite{BM, MTsection}. 
\par We particularly discuss bisection geometries given by equations of the following form, as studied in \cite{Kimura201603}:
\begin{equation}
\label{double cover K3 general in 3}
u^2=\prod^4_{i=1} (t-\alpha_i)\, x^4+\prod^8_{j=5} (t-\alpha_j),
\end{equation}
where $t$ and $x$ represent inhomogeneous coordinates of $\P^1$, and each $\alpha_i$, $i=1, \ldots, 8$, is a point in $\P^1$. Thus, the equation (\ref{double cover K3 general in 3}) yields a double cover of $\P^1\times\P^1$ branched over a (4,4) curve, which is a K3 surface \cite{Kimura201603}. A projection onto $\P^1$ in the product $\P^1\times\P^1$ yields a genus-one fibration, as discussed in \cite{Kimura201603}. We regard $t$ as the coordinate of the base $\P^1$ of the genus-one fibration. The K3 surface (\ref{double cover K3 general in 3}) is a bisection geometry \cite{Kimura201603}. The elliptic fibers of this fibration have a particular symmetry, which forces the complex structure of the fibers to be constant with $\tau=i$ \footnote{Equivalently, they have j-invariant 1728.} throughout the base $\P^1$ \cite{Kimura201603}. This constrains the types of 7-branes so that the fiber types over the 7-branes can only be $III$, $I_0^*$, or $III^*$ \cite{Kimura201603}.
\par We focus on the case of $\alpha_5=\alpha_6=0$ here. Then, the equation of the K3 surface (\ref{double cover K3 general in 3}) becomes 
\begin{equation}
\label{K3 special in 3}
u^2=\prod^4_{i=1} (t-\alpha_i)\, x^4+t^2 (t-\alpha_7)(t-\alpha_8).
\end{equation}
We only consider the cases where $\alpha_i$, $i=1,2,3,4$, are mutually distinct and $\alpha_i$, $i=1,2,3,4,7,8$ are non-zero. The discriminant of the K3 surface (\ref{K3 special in 3}) is given by \cite{Kimura201603}
\begin{equation}
\Delta \sim \prod^4_{i=1} (t-\alpha_i)^3 \cdot t^6 (t-\alpha_7)^3(t-\alpha_8)^3.
\end{equation}
7-branes at the origin $t=0$ have type $D_4$, and each stack of 7-branes at $t=\alpha_i$, $i=1,2,3,4$, has type $H_1$ \cite{Kimura201603}. 7-brane stacks at $t=\alpha_7, \alpha_8$ have type $H_1$ when $\alpha_7$ and $\alpha_8$ are not equal. At the limit at which $\alpha_7$ and $\alpha_8$ coincide, the 7-brane type at $t=\alpha_7$ is enhanced to $D_4$. 
\par For the reader's convenience, we present the Jacobian fibration of the double cover (\ref{K3 special in 3}). The Jacobian fibration admits transformation to the following Weierstrass form \cite{Muk}:
\begin{equation}
y^2=\frac{1}{4}x^3-\prod^4_{i=1} (t-\alpha_i)\, t^2 (t-\alpha_7)(t-\alpha_8)x.
\end{equation}
\par We consider F-theory on a K3 surface (\ref{K3 special in 3}) times a K3 surface. We consider the situation where the probe D3-brane is placed at the origin $t=0$. We consider tilting the stack of 7-branes at the origin $t=0$ to yield a 4d $N=1$ SCFT, as discussed in section \ref{sec2}. For general parameters $\alpha_i$, $i=1,2,3,4,7,8$, the K3 surface (\ref{K3 special in 3}) has a bisection, but it does not have a global section, and a $\Z_2$ gauge group forms in F-theory. As we discussed previously, this is not probed by the D3-brane in IR. 
\par One can tune the parameters so a bisection splits into two global sections. A bisection splits into global sections when either the coefficient of $x^4$ or constant term in the quartic polynomial becomes a perfect square \cite{MTsection}. This occurs when $\alpha_7$ and $\alpha_8$ become coincident in (\ref{K3 special in 3}). Therefore, at the limit at which $\alpha_7$ and $\alpha_8$ coincide, the bisection splits into two global sections. Then, U(1), instead of the $\Z_2$ gauge group, forms in F-theory, but the theory on the D3-brane probe is insensitive to this change. 
\par One can choose $\alpha_7$ and $\alpha_8$ far away from the origin so they do not lie in the local patch that the D3-brane probes. Tuning $\alpha_8$ such that it approaches $\alpha_7$  does not affect the singularity type at the origin $t=0$. 
\par We particularly considered a bisection given by a specific form of an equation, but a similar reasoning applies to general bisection geometry. 

\section{Puzzle in the global structure of the gauge group}
\label{sec4}
\par When an elliptic fibration has a section, the set of sections is known to form a group, and this group is referred to as the ``Mordell--Weil group.'' When the Mordell--Weil group of an elliptic fibration has a torsion part, the torsion part divides the global structure of the gauge group formed in F-theory on that elliptic fibration \cite{AspinwallGross, AMrational, MMTW}. 
\par There are non-isomorphic elliptic K3 surfaces whose singular fiber types are identical and F-theory compactifications yield identical 7-brane types but their Mordell--Weil torsions are different. We explicitly give an example of this. We utilize extremal K3 surfaces \footnote{Discussions of extremal elliptic K3 surfaces in the context of string theory can be found, e.g., in \cite{Kimura2015, Kimura201603, Kimura201607, Kimura201810, Kimura201902, Font2020}.}. A complex elliptically fibered K3 surface $f: S \rightarrow \P^1$ with a global section is referred to as extremal when the Picard number of K3 surface $S$ is 20 and the Mordell--Weil group, $MW(S,f)$, is a finite group. Complex K3 surfaces with the Picard number of 20 are referred to as attractive K3 surfaces \footnote{We follow the convention of the term used in \cite{Moore}.}.  
\par It is a mathematical fact that the complex structures of the attractive K3 surfaces are labelled by triplets of integers \cite{PS-S, SI}. The complex structures, singularity types, and Mordell--Weil torsions of the K3 extremal fibrations were completely classified in \cite{ShimadaZhang}. A pair of non-isomorphic K3 surfaces whose singularity types are identical but Mordell--Weil torsions are different can be found in Table 2 in \cite{ShimadaZhang}; numbers are assigned to the singularity types of the K3 extremal fibrations in this table. No. 276 in this table corresponds to the singularity type $E_7A_9A_2$. There are two K3 surfaces possessing this singularity type according to the table, and they have different complex structures, so they are non-isomorphic. F-theory on either of these two K3 surfaces times a K3 surface yields a 4d $N=2$ theory, and an $E_7\times SU(10)\times SU(3)$ gauge group forms in both compactifications. However, one of the extremal K3 surfaces has no Mordell--Weil torsion, while the other K3 surface has Mordell--Weil torsion isomorphic to $\Z_2$, so the gauge groups formed in the two theories are globally different. These theories \footnote{We chose an example of extremal K3 fibrations, whose classification is known. However, the appearance of two non-isomorphic K3 surfaces whose singularity types are identical while having distinct Mordell--Weil torsions is not limited to extremal K3 surfaces, and such cases typically arise in the whole moduli of algebraic K3 surfaces.} yield different theories in UV, possessing distinct global structures of the gauge groups.  
\par A natural question is whether the Mordell--Weil torsional group dividing the gauge group is reflected in the theory on the D3-brane probe, which probes only a local patch. Let us take a local patch of an elliptic fibration and place a probe D3-brane in the neighborhood of a stack of 7-branes. Locally, one cannot see the structures of global sections; thus, the Mordell--Weil group cannot be seen locally. Then, it is natural to consider that the Mordell--Weil torsions also cannot be locally seen. A consequence of this reasoning seems to be that the effect of Mordell--Weil torsion is not reflected in the 4d $N=1$ SCFT on the probe D3-brane in IR. 
\par However, the 7-brane type is determined by the local information. In the language of F-theory, the 7-brane type is determined by the type of a singular fiber over the 7-branes, which is ``vertical'' information not relevant to the global feature of the base space. Because the Mordell--Weil torsional group acts on the gauge groups, the effect of this group might be reflected in the flavor symmetry group formed on the 7-branes that the D3-brane probes.

\par Because these two viewpoints provide opposing interpretations, one must choose one of the two interpretations. The question of which one gives the correct physical interpretation is left for future study. 

\section{Concluding remarks and open problem}
\label{sec5}
\par In this note, we analyzed 4d $N=1$ SCFTs obtained as deformations of 4d $N=2$ SCFTs on S-folds by tilting 7-branes. We showed that T-branes receive constraints from the geometry when S-fold constructions are considered. Consequently, brane monodromies are also constrained by the geometric conditions, and we presented explicit examples of this. 
\par We also discussed two 4d $N=1$ SCFTs behaving identically in IR but flowing to distinct theories in UV. To construct these theories, utilizing the global structure of genus-one fibrations lacking a global section was useful. Our argument also showed that U(1) and discrete $\Z_n$ groups arising from the global feature of the geometry in F-theory are not reflected in 4d $N=1$ SCFTs on the probe D3-brane. 
\par We encountered a dilemma when considering an effect of the global geometry on the theory living on the probe D3-brane, i.e., the action of the Mordell--Weil torsional group on the flavor symmetry group. Two opposing viewpoints seem possible for this problem, and which viewpoint yields a correct interpretation is left to be determined in future work.
\par We utilized the property that the D3-brane only probes a local patch of the geometry to deduce several physical consequences in sections \ref{sec3} and \ref{sec4}. Strictly speaking, however, when the space possesses an S-fold singularity, not only the local patch of the geometry and the modes coming from the interaction of the D3-branes and the S-fold but also the structure of the S-fold can contribute to the 4d physics, as explained in \cite{Garcia-Etxebarria2015}. It might be interesting to investigate this subtle point, and this is left for future studies.
\par We remark that the conformal anomalies $a_{IR}$ and $c_{IR}$ for 4d $N=1$ SCFTs analyzed in this study can be computed. The value of the trial central charge, $a_{IR}(t)$, is expressed in terms of $a_{UV}$, $c_{UV}$, $k_{UV}$ and a parameter of the Jordan block structure pertaining to the superpotential deformation $\delta W=Tr_G(\phi(Z_1, Z_2)\cdot \mathcal{O})$ utilizing 't Hooft anomaly matching \cite{Heckman201009}. The conformal anomalies $a_{IR}$ and $c_{IR}$ are obtained by applying the method in \cite{Heckman201009}, or the method in \cite{Apruzzi201808} with the geometric interpretation given in \cite{Carta201809}.

\section*{Acknowledgments}

We would like to thank Yosuke Imamura, Shun'ya Mizoguchi and Shigeru Mukai for discussions. We are also grateful to the referee for improving this manuscript.


\begin{thebibliography}{99}

\bibitem{SeibergWitten199407}
N.~Seiberg and E.~Witten,
``Electric - magnetic duality, monopole condensation, and confinement in N=2 supersymmetric Yang-Mills theory,''
{\it Nucl. Phys.} \textbf{B426}, 19--52 (1994)
[erratum: {\it Nucl. Phys.} \textbf{B430}, 485--486 (1994)]
[arXiv:hep-th/9407087 [hep-th]].

\bibitem{SeibergWitten199408}
N.~Seiberg and E.~Witten,
``Monopoles, duality and chiral symmetry breaking in N=2 supersymmetric QCD,''
{\it Nucl. Phys.} \textbf{B431}, 484--550 (1994)
[arXiv:hep-th/9408099 [hep-th]].

\bibitem{Argyres1994}
P.~C.~Argyres and A.~E.~Faraggi,
``The vacuum structure and spectrum of N=2 supersymmetric SU(n) gauge theory,''
{\it Phys. Rev. Lett.} \textbf{74}, 3931--3934 (1995)
[arXiv:hep-th/9411057 [hep-th]].

\bibitem{Argyres1995}
P.~C.~Argyres and M.~R.~Douglas,
``New phenomena in SU(3) supersymmetric gauge theory,''
{\it Nucl. Phys.} \textbf{B448}, 93--126 (1995)
[arXiv:hep-th/9505062 [hep-th]].

\bibitem{Argyres199511}
P.~C.~Argyres, M.~R.~Plesser, N.~Seiberg and E.~Witten,
``New N=2 superconformal field theories in four-dimensions,''
{\it Nucl. Phys.} \textbf{B461}, 71--84 (1996)
[arXiv:hep-th/9511154 [hep-th]].

\bibitem{Eguchi1996}
T.~Eguchi, K.~Hori, K.~Ito and S.~K.~Yang,
``Study of N=2 superconformal field theories in four-dimensions,''
{\it Nucl. Phys.} \textbf{B471}, 430--444 (1996)
[arXiv:hep-th/9603002 [hep-th]].

\bibitem{Minahan199608}
J.~A.~Minahan and D.~Nemeschansky,
``An N=2 superconformal fixed point with E(6) global symmetry,''
{\it Nucl. Phys.} \textbf{B482}, 142--152 (1996)
[arXiv:hep-th/9608047 [hep-th]].

\bibitem{Minahan199610}
J.~A.~Minahan and D.~Nemeschansky,
``Superconformal fixed points with E(n) global symmetry,''
{\it Nucl. Phys.} \textbf{B489}, 24--46 (1997)
[arXiv:hep-th/9610076 [hep-th]].

\bibitem{Argyres2007}
P.~C.~Argyres and J.~R.~Wittig,
``Infinite coupling duals of N=2 gauge theories and new rank 1 superconformal field theories,''
{\it JHEP} \textbf{01}, 074 (2008)
[arXiv:0712.2028 [hep-th]].

\bibitem{Gaiotto2009}
D.~Gaiotto,
``N=2 dualities,''
{\it JHEP} \textbf{08}, 034 (2012)
[arXiv:0904.2715 [hep-th]].

\bibitem{Cecotti2010}
S.~Cecotti, A.~Neitzke and C.~Vafa,
``R-Twisting and 4d/2d Correspondences,''
[arXiv:1006.3435 [hep-th]].

\bibitem{Argyres2015}
P.~Argyres, M.~Lotito, Y.~L\"u and M.~Martone,
``Geometric constraints on the space of $ \mathcal{N} $ = 2 SCFTs. Part I: physical constraints on relevant deformations,''
{\it JHEP} \textbf{02}, 001 (2018)
[arXiv:1505.04814 [hep-th]].

\bibitem{Xie2015}
D.~Xie and S.~T.~Yau,
``4d N=2 SCFT and singularity theory Part I: Classification,''
[arXiv:1510.01324 [hep-th]].

\bibitem{Argyres201611}
P.~C.~Argyres and M.~Martone,
``4d $ \mathcal{N} $ =2 theories with disconnected gauge groups,''
{\it JHEP} \textbf{03}, 145 (2017)
[arXiv:1611.08602 [hep-th]].

\bibitem{Caorsi2018}
M.~Caorsi and S.~Cecotti,
``Geometric classification of 4d $\mathcal{N}=2$ SCFTs,''
{\it JHEP} \textbf{07}, 138 (2018)
[arXiv:1801.04542 [hep-th]].

\bibitem{Borsten2018}
L.~Borsten, M.~J.~Duff and A.~Marrani,
``Twin conformal field theories,''
{\it JHEP} \textbf{03}, 112 (2019)
[arXiv:1812.11130 [hep-th]].

\bibitem{Apruzzi2020}
F.~Apruzzi, S.~Giacomelli and S.~Sch\"afer-Nameki,
``4d $\mathcal{N}=2$ S-folds,''
{\it Phys. Rev.} \textbf{D101}, no.10, 106008 (2020)
[arXiv:2001.00533 [hep-th]].

\bibitem{Argyres202003}
P.~Argyres and M.~Martone,
``Construction and classification of Coulomb branch geometries,''
[arXiv:2003.04954 [hep-th]].

\bibitem{He202004}
Y.~H.~He, E.~Hirst and T.~Peterken,
``Machine-Learning Dessins d'Enfants: Explorations via Modular and Seiberg-Witten Curves,''
[arXiv:2004.05218 [hep-th]].

\bibitem{Bourget2020}
A.~Bourget, J.~F.~Grimminger, A.~Hanany, M.~Sperling, G.~Zafrir and Z.~Zhong,
``Magnetic quivers for rank 1 theories,''
{\it JHEP} \textbf{09}, 189 (2020)
[arXiv:2006.16994 [hep-th]].

\bibitem{Giacomelli202007}
S.~Giacomelli, C.~Meneghelli and W.~Peelaers,
``New N=2 superconformal field theories from S-folds,''
[arXiv:2007.00647 [hep-th]].

\bibitem{Heckman2020}
J.~J.~Heckman, C.~Lawrie, T.~B.~Rochais, H.~Y.~Zhang and G.~Zoccarato,
``S-folds, String Junctions, and 4D $\mathcal{N} = 2$ SCFTs,''
[arXiv:2009.10090 [hep-th]].

\bibitem{Giacomelli202010}
S.~Giacomelli, M.~Martone, Y.~Tachikawa and G.~Zafrir,
``More on $\mathcal{N} =2$ S-folds,''
[arXiv:2010.03943 [hep-th]].

\bibitem{Garcia-Etxebarria2015}
I.~Garc\'\i{}a-Etxebarria and D.~Regalado,
``$ \mathcal{N}=3 $ four dimensional field theories,''
{\it JHEP} \textbf{03}, 083 (2016)
[arXiv:1512.06434 [hep-th]].

\bibitem{Aharony2015}
O.~Aharony and M.~Evtikhiev,
``On four dimensional N = 3 superconformal theories,''
{\it JHEP} \textbf{04}, 040 (2016)
[arXiv:1512.03524 [hep-th]].

\bibitem{Nishinaka2016}
T.~Nishinaka and Y.~Tachikawa,
``On 4d rank-one $ \mathcal{N}=3 $ superconformal field theories,''
{\it JHEP} \textbf{09}, 116 (2016)
[arXiv:1602.01503 [hep-th]].

\bibitem{Aharony2016}
O.~Aharony and Y.~Tachikawa,
``S-folds and 4d N=3 superconformal field theories,''
{\it JHEP} \textbf{06}, 044 (2016)
[arXiv:1602.08638 [hep-th]].

\bibitem{Imamura2016}
Y.~Imamura and S.~Yokoyama,
``Superconformal index of ${ \mathcal N }=3$ orientifold theories,''
{\it J. Phys.} \textbf{A49}, no.43, 435401 (2016)
[arXiv:1603.00851 [hep-th]].

\bibitem{Imamura201606}
Y.~Imamura, H.~Kato and D.~Yokoyama,
``Supersymmetry Enhancement and Junctions in S-folds,''
{\it JHEP} \textbf{10}, 150 (2016)
[arXiv:1606.07186 [hep-th]].

\bibitem{Agarwal2016}
P.~Agarwal and A.~Amariti,
``Notes on S-folds and $ \mathcal{N} $ = 3 theories,''
{\it JHEP} \textbf{09}, 032 (2016)
[arXiv:1607.00313 [hep-th]].

\bibitem{Lemos2016}
M.~Lemos, P.~Liendo, C.~Meneghelli and V.~Mitev,
``Bootstrapping $\mathcal{N}=3$ superconformal theories,''
{\it JHEP} \textbf{04}, 032 (2017)
[arXiv:1612.01536 [hep-th]].

\bibitem{vanMuiden2017}
J.~van Muiden and A.~Van Proeyen,
``The $ \mathcal{N} $ = 3 Weyl multiplet in four dimensions,''
{\it JHEP} \textbf{01}, 167 (2019)
[arXiv:1702.06442 [hep-th]].

\bibitem{Bourton2018}
T.~Bourton, A.~Pini and E.~Pomoni,
``4d $\mathcal{N}=3$ indices via discrete gauging,''
{\it JHEP} \textbf{10}, 131 (2018)
[arXiv:1804.05396 [hep-th]].

\bibitem{Arai2018}
R.~Arai, S.~Fujiwara and Y.~Imamura,
``BPS Partition Functions for S-folds,''
{\it JHEP} \textbf{03}, 172 (2019)
[arXiv:1901.00023 [hep-th]].

\bibitem{CCHV2010}
S.~Cecotti, C.~Cordova, J.~J.~Heckman and C.~Vafa,
``T-Branes and Monodromy,''
{\it JHEP} \textbf{07} (2011) 030 
[arXiv:1010.5780 [hep-th]].

\bibitem{DKS2003}
R.~Donagi, S.~Katz and E.~Sharpe,
``Spectra of D-branes with higgs vevs,''
{\it Adv. Theor. Math. Phys.} \textbf{8} (2004) no.5, 813--859 
[arXiv:hep-th/0309270].

\bibitem{DW2011}
R.~Donagi and M.~Wijnholt,
``Gluing Branes, I,''
{\it JHEP} \textbf{05} (2013) 068 
[arXiv:1104.2610 [hep-th]].

\bibitem{MTmatter}D.~R.~Morrison and W.~Taylor, ``Matter and singularities,'' {\it JHEP} {\bf 01} (2012) 022 [arXiv:1106.3563 [hep-th]]. 

\bibitem{MT1204}D.~R.~Morrison and W.~Taylor, ``Toric bases for 6D F-theory models,'' {\it Fortsch. Phys.} {\bf 60} (2012) 1187--1216 [arXiv:1204.0283 [hep-th]].

\bibitem{AHK2013}
L.~B.~Anderson, J.~J.~Heckman and S.~Katz,
``T-Branes and Geometry,''
{\it JHEP} \textbf{05} (2014) 080 
[arXiv:1310.1931 [hep-th]].

\bibitem{CR2014}
A.~Collinucci and R.~Savelli,
``T-branes as branes within branes,''
{\it JHEP} \textbf{09} (2015) 161 
[arXiv:1410.4178 [hep-th]].

\bibitem{CR2014space}
A.~Collinucci and R.~Savelli,
``F-theory on singular spaces,''
{\it JHEP} \textbf{09} (2015) 100 
[arXiv:1410.4867 [hep-th]].

\bibitem{CQV2015}
M.~Cicoli, F.~Quevedo and R.~Valandro,
``De Sitter from T-branes,''
{\it JHEP} \textbf{03} (2016) 141 
[arXiv:1512.04558 [hep-th]].

\bibitem{HHRRZ1907}
F.~Hassler, J.~J.~Heckman, T.~B.~Rochais, T.~Rudelius and H.~Y.~Zhang,
``T-Branes, String Junctions, and 6D SCFTs,''
{\it Phys. Rev.} \textbf{D101} (2020) no.8, 086018 
[arXiv:1907.11230 [hep-th]].

\bibitem{Heckman201006}
J.~J.~Heckman and C.~Vafa,
``An Exceptional Sector for F-theory GUTs,''
{\it Phys. Rev.} \textbf{D83}, 026006 (2011)
[arXiv:1006.5459 [hep-th]].

\bibitem{Heckman201009}
J.~J.~Heckman, Y.~Tachikawa, C.~Vafa and B.~Wecht,
``N = 1 SCFTs from Brane Monodromy,''
{\it JHEP} \textbf{11}, 132 (2010)
[arXiv:1009.0017 [hep-th]].

\bibitem{Banks1996}
T.~Banks, M.~R.~Douglas and N.~Seiberg,
``Probing F theory with branes,''
{\it Phys. Lett.} \textbf{B387}, 278-281 (1996)
[arXiv:hep-th/9605199 [hep-th]].

\bibitem{Douglas1996}
M.~R.~Douglas, D.~A.~Lowe and J.~H.~Schwarz,
``Probing F theory with multiple branes,''
{\it Phys. Lett.} \textbf{B394}, 297-301 (1997)
[arXiv:hep-th/9612062 [hep-th]].

\bibitem{Fayyazuddin1998}
A.~Fayyazuddin and M.~Spalinski,
``Large N superconformal gauge theories and supergravity orientifolds,''
{\it Nucl. Phys.} \textbf{B535}, 219--232 (1998)
[arXiv:hep-th/9805096 [hep-th]].

\bibitem{Aharony199806}
O.~Aharony, A.~Fayyazuddin and J.~M.~Maldacena,
``The Large N limit of N=2, N=1 field theories from three-branes in F theory,''
{\it JHEP} \textbf{07}, 013 (1998)
[arXiv:hep-th/9806159 [hep-th]].

\bibitem{Gukov199808}
S.~Gukov and A.~Kapustin,
``New N=2 superconformal field theories from M / F theory orbifolds,''
{\it Nucl. Phys.} \textbf{B545}, 283--308 (1999)
[arXiv:hep-th/9808175 [hep-th]].

\bibitem{Bah2011}
I.~Bah and B.~Wecht,
``New N=1 Superconformal Field Theories In Four Dimensions,''
{\it JHEP} \textbf{07}, 107 (2013)
[arXiv:1111.3402 [hep-th]].

\bibitem{Gadde2013}
A.~Gadde, K.~Maruyoshi, Y.~Tachikawa and W.~Yan,
``New N=1 Dualities,''
{\it JHEP} \textbf{06}, 056 (2013)
[arXiv:1303.0836 [hep-th]].

\bibitem{McGrane2014}
J.~McGrane and B.~Wecht,
``Theories of class $ \mathcal{S} $ and new $ \mathcal{N} $ = 1 SCFTs,''
{\it JHEP} \textbf{06}, 047 (2015)
[arXiv:1409.7668 [hep-th]].

\bibitem{Maruyoshi2016}
K.~Maruyoshi and J.~Song,
``$ \mathcal{N}=1 $ deformations and RG flows of $ \mathcal{N}=2 $ SCFTs,''
{\it JHEP} \textbf{02}, 075 (2017)
[arXiv:1607.04281 [hep-th]].

\bibitem{Bourton2020}
T.~Bourton, A.~Pini and E.~Pomoni,
``The Coulomb and Higgs Branches of $\mathcal{N}=1$ Theories of Class $\mathcal{S}_k$,''
[arXiv:2011.01587 [hep-th]].

\bibitem{Hayashi2009}
H.~Hayashi, T.~Kawano, R.~Tatar and T.~Watari,
``Codimension-3 Singularities and Yukawa Couplings in F-theory,''
{\it Nucl. Phys.} \textbf{B823}, 47--115 (2009)
[arXiv:0901.4941 [hep-th]].

\bibitem{Donagi200904}
R.~Donagi and M.~Wijnholt,
``Higgs Bundles and UV Completion in F-Theory,''
{\it Commun. Math. Phys.} \textbf{326}, 287--327 (2014)
[arXiv:0904.1218 [hep-th]].

\bibitem{Bouchard2009}
V.~Bouchard, J.~J.~Heckman, J.~Seo and C.~Vafa,
``F-theory and Neutrinos: Kaluza-Klein Dilution of Flavor Hierarchy,''
{\it JHEP} \textbf{01}, 061 (2010)
[arXiv:0904.1419 [hep-ph]].

\bibitem{Heckman200906}
J.~J.~Heckman, A.~Tavanfar and C.~Vafa,
``The Point of E(8) in F-theory GUTs,''
{\it JHEP} \textbf{08}, 040 (2010)
[arXiv:0906.0581 [hep-th]].

\bibitem{Marsano2009}
J.~Marsano, N.~Saulina and S.~Schafer-Nameki,
``Monodromies, Fluxes, and Compact Three-Generation F-theory GUTs,''
{\it JHEP} \textbf{08}, 046 (2009)
[arXiv:0906.4672 [hep-th]].

\bibitem{Dasgupta1996}
K.~Dasgupta and S.~Mukhi,
``F theory at constant coupling,''
{\it Phys. Lett.} \textbf{B385}, 125--131 (1996)
[arXiv:hep-th/9606044 [hep-th]].

\bibitem{Kimura2015}Y.~Kimura, ``Gauge Groups and Matter Fields on Some Models of F-theory without Section'', {\it JHEP} {\bf 03} (2016) 042 [arXiv:1511.06912 [hep-th]].
\bibitem{Kimura201603}Y.~Kimura, ``Gauge symmetries and matter fields in F-theory models without section- compactifications on double cover and Fermat quartic K3 constructions times K3'', {\it Adv. Theor. Math. Phys.} {\bf 21} (2017) no.8, 2087--2114 [arXiv:1603.03212 [hep-th]].
\bibitem{Kimura201607}Y.~Kimura, ``Gauge groups and matter spectra in F-theory compactifications on genus-one fibered Calabi-Yau 4-folds without section - Hypersurface and double cover constructions'', {\it Adv. Theor. Math. Phys.} {\bf 22} (2018) no.6, 1489--1533 [arXiv:1607.02978 [hep-th]]. 
\bibitem{Kimura201810}Y.~Kimura, ``Nongeometric heterotic strings and dual F-theory with enhanced gauge groups'', {\it JHEP} {\bf 02} (2019) 036 [arXiv:1810.07657 [hep-th]].
\bibitem{Kimura201902}Y.~Kimura, ``Unbroken $E_7\times E_7$ nongeometric heterotic strings, stable degenerations and enhanced gauge groups in F-theory duals'' [arXiv:1902.00944 [hep-th]].

\bibitem{BM}V.~Braun and D.~R.~Morrison, ``F-theory on Genus-One Fibrations'', {\it JHEP} {\bf 08} (2014) 132 [arXiv:1401.7844 [hep-th]].
\bibitem{MTsection}D.~R.~Morrison and W.~Taylor, ``Sections, multisections, and $U(1)$ fields in F-theory'', {\it J. Singularities} {\bf 15} (2016) 126--149 [arXiv:1404.1527 [hep-th]].
\bibitem{AGGK}L.~B.~Anderson, I.~Garcia-Etxebarria, T.~W.~Grimm and J.~Keitel, ``Physics of F-theory compactifications without section'', {\it JHEP} {\bf 12} (2014) 156 [arXiv:1406.5180 [hep-th]].
\bibitem{KMOPR}D.~Klevers, D.~K.~Mayorga Pena, P.~K.~Oehlmann, H.~Piragua and J.~Reuter, ``F-Theory on all Toric Hypersurface Fibrations and its Higgs Branches'', {\it JHEP} {\bf 01} (2015) 142 [arXiv:1408.4808 [hep-th]].
\bibitem{GGK}I.~Garcia-Etxebarria, T.~W.~Grimm and J.~Keitel, ``Yukawas and discrete symmetries in F-theory compactifications without section'', {\it JHEP} {\bf 11} (2014) 125 [arXiv:1408.6448 [hep-th]].
\bibitem{MPTW}C.~Mayrhofer, E.~Palti, O.~Till and T.~Weigand, ``Discrete Gauge Symmetries by Higgsing in four-dimensional F-Theory Compactifications'', {\it JHEP} {\bf 12} (2014) 068 [arXiv:1408.6831 [hep-th]].
\bibitem{MPTW2}C.~Mayrhofer, E.~Palti, O.~Till and T.~Weigand, ``On Discrete Symmetries and Torsion Homology in F-Theory'', {\it JHEP} {\bf 06} (2015) 029 [arXiv:1410.7814 [hep-th]].
\bibitem{BGKintfiber}V.~Braun, T.~W.~Grimm and J.~Keitel, ``Complete Intersection Fibers in F-Theory'', {\it JHEP} {\bf 03} (2015) 125 [arXiv:1411.2615 [hep-th]].
\bibitem{CDKPP}M.~Cveti\v c, R.~Donagi, D.~Klevers, H.~Piragua and M.~Poretschkin, ``F-theory vacua with $\mathbb Z_3$ gauge symmetry'', {\it Nucl. Phys.} {\bf B898} (2015) 736--750 [arXiv:1502.06953 [hep-th]].
\bibitem{LMTW}L.~Lin, C.~Mayrhofer, O.~Till and T.~Weigand, ``Fluxes in F-theory Compactifications on Genus-One Fibrations'', {\it JHEP} {\bf 01} (2016) 098 [arXiv:1508.00162 [hep-th]].
\bibitem{Kimura201608}Y.~Kimura, ``Discrete Gauge Groups in F-theory Models on Genus-One Fibered Calabi-Yau 4-folds without Section'', {\it JHEP} {\bf 04} (2017) 168 [arXiv:1608.07219 [hep-th]].
\bibitem{Kimura201905}Y.~Kimura, ``Discrete gauge groups in certain F-theory models in six dimensions'', {\it JHEP} {\bf 07} (2019) 027 [arXiv:1905.03775 [hep-th]].

\bibitem{Kimura201907}
Y.~Kimura,
``A note on transition in discrete gauge groups in F-theory,''
{\it Int. J. Mod. Phys.} \textbf{A35}, no.24, 2050144 (2020)
[arXiv:1907.13503 [hep-th]].

\bibitem{Kimura201908}
Y.~Kimura,
``F-theory models with $U(1)\times \mathbb{Z}_2,\, \mathbb{Z}_4$ and transitions in discrete gauge groups,''
{\it JHEP} \textbf{03} (2020) 153 
[arXiv:1908.06621 [hep-th]].

\bibitem{MV2}D.~R.~Morrison and C.~Vafa, ``Compactifications of F-theory on Calabi-Yau threefolds. 2'', {\it Nucl. Phys.} {\bf B 476} (1996) 437 [arXiv:hep-th/9603161].
\bibitem{Vaf}C.~Vafa, ``Evidence for F-theory'', {\it Nucl. Phys.} {\bf B 469} (1996) 403 [arXiv:hep-th/9602022].
\bibitem{MV1}D.~R.~Morrison and C.~Vafa, ``Compactifications of F-theory on Calabi-Yau threefolds. 1'', {\it Nucl. Phys.} {\bf B 473} (1996) 74 [arXiv:hep-th/9602114].

\bibitem{DWmodel}R.~Donagi and M.~Wijnholt, ``Model Building with F-Theory,'' {\it Adv. Theor. Math. Phys.}  {\bf 15}, 1237 (2011) [arXiv:0802.2969 [hep-th]].
\bibitem{BHV}C.~Beasley, J.~J.~Heckman and C.~Vafa, ``GUTs and Exceptional Branes in F-theory - I,'' {\it JHEP} {\bf 0901}, 058 (2009) [arXiv:0802.3391 [hep-th]].
\bibitem{BHV2}C.~Beasley, J.~J.~Heckman and C.~Vafa, ``GUTs and Exceptional Branes in F-theory - II: Experimental Predictions,'' {\it JHEP} {\bf 0901}, 059 (2009) [arXiv:0806.0102 [hep-th]].
\bibitem{DWGUT}R.~Donagi and M.~Wijnholt, ``Breaking GUT Groups in F-Theory,'' {\it Adv. Theor. Math. Phys.}  {\bf 15}, 1523 (2011) [arXiv:0808.2223 [hep-th]].

\bibitem{MorrisonPark}D.~R.~Morrison and D.~S.~Park, ``F-Theory and the Mordell-Weil Group of Elliptically-Fibered Calabi-Yau Threefolds'', {\it JHEP} {\bf 10} (2012) 128 [arXiv:1208.2695 [hep-th]].
\bibitem{MPW}C.~Mayrhofer, E.~Palti and T.~Weigand, ``U(1) symmetries in F-theory GUTs with multiple sections'', {\it JHEP} {\bf 03} (2013) 098 [arXiv:1211.6742 [hep-th]].
\bibitem{BMPWsection}J.~Borchmann, C.~Mayrhofer, E.~Palti and T.~Weigand, ``Elliptic fibrations for $SU(5)\times U(1)\times U(1)$ F-theory vacua'', {\it Phys. Rev.} {\bf D88} (2013) no.4 046005 [arXiv:1303.5054 [hep-th]].
\bibitem{AL}I.~Antoniadis and G.~K.~Leontaris, ``F-GUTs with Mordell-Weil U(1)'s,'' {\it Phys. Lett.} {\bf B735} (2014) 226--230 [arXiv:1404.6720 [hep-th]].
\bibitem{MM}A.~Malmendier and D.~R.~Morrison, ``K3 surfaces, modular forms, and non-geometric heterotic compactifications'', {\it Lett. Math. Phys.} {\bf 105} (2015) no.8, 1085--1118 [arXiv:1406.4873 [hep-th]].
\bibitem{LSW}C.~Lawrie, S.~Sch\"afer-Nameki and J.-M.~Wong, ``F-theory and All Things Rational: Surveying U(1) Symmetries with Rational Sections'', {\it JHEP} {\bf 09} (2015) 144 [arXiv:1504.05593 [hep-th]]. 
\bibitem{CKPT}M.~Cveti\v c, D.~Klevers, H.~Piragua and W.~Taylor, ``General U(1)$\times$U(1) F-theory compactifications and beyond: geometry of unHiggsings and novel matter structure,'' {\it JHEP} {\bf 1511} (2015) 204 [arXiv:1507.05954 [hep-th]].
\bibitem{MPT1610}D.~R.~Morrison, D.~S.~Park and W.~Taylor, ``Non-Higgsable abelian gauge symmetry and $\mathrm{F}$-theory on fiber products of rational elliptic surfaces'', {\it Adv. Theor. Math. Phys.} {\bf 22} (2018) 177--245 [arXiv:1610.06929 [hep-th]].
\bibitem{Kimura1802}Y.~Kimura, ``F-theory models on K3 surfaces with various Mordell-Weil ranks -constructions that use quadratic base change of rational elliptic surfaces'', {\it JHEP} {\bf 05} (2018) 048 [arXiv:1802.05195 [hep-th]].
\bibitem{MizTani2018}S.~Mizoguchi and T.~Tani, ``Non-Cartan Mordell-Weil lattices of rational elliptic surfaces and heterotic/F-theory compactifications'', {\it JHEP} {\bf 03} (2019) 121 [arXiv:1808.08001 [hep-th]].
\bibitem{LW1905}S.-J.~Lee and T.~Weigand, ``Swampland Bounds on the Abelian Gauge Sector'', {\it Phys. Rev.} {\bf D100} (2019) no.2 026015 [arXiv:1905.13213 [hep-th]].
\bibitem{Kimura1910}Y.~Kimura, ``$\frac{1}{2}$ Calabi-Yau 3-folds, Calabi-Yau 3-folds as double covers, and F-theory with U(1)s'', {\it JHEP} {\bf 02} (2020) 076 [arXiv:1910.00008 [hep-th]].
\bibitem{CKS1910}C.~F.~Cota, A.~Klemm, and T.~Schimannek, ``Topological strings on genus one fibered Calabi-Yau 3-folds and string dualities'', {\it JHEP} {\bf 11} (2019) 170 [arXiv:1910.01988 [hep-th]].

\bibitem{Chabrol2019}
L.~Chabrol,
``F-theory and Heterotic Duality, Weierstrass Models from Wilson lines,''
{\it Eur. Phys. J.} \textbf{C80}, no.10, 944 (2020)
[arXiv:1910.12844 [hep-th]].

\bibitem{Kimura1911}Y.~Kimura, ``$\frac{1}{2}$Calabi-Yau 4-folds and four-dimensional F-theory on Calabi-Yau 4-folds with U(1) factors'' [arXiv:1911.03960 [hep-th]].
\bibitem{FKKMT1912}S.~Fukuchi, N.~Kan, R.~Kuramochi, S.~Mizoguchi and H.~Tashiro, ``More on a dessin on the base: Kodaira exceptional fibers and mutually (non-)local branes'', {\it Phys.Lett.} {\bf B803} (2020) 135333 [arXiv:1912.02974 [hep-th]].

\bibitem{Kan2020}
N.~Kan, S.~Mizoguchi and T.~Tani,
``Half-hypermultiplets and incomplete/complete resolutions in F-theory,''
{\it JHEP} \textbf{08}, 063 (2020)
[arXiv:2003.05563 [hep-th]].

\bibitem{Karozas2020}
A.~Karozas, G.~K.~Leontaris, I.~Tavellaris and N.~D.~Vlachos,
``On the LHC signatures of $SU(5)\times U(1)'$ F-theory motivated models,''
[arXiv:2007.05936 [hep-ph]].

\bibitem{Angelantonj2020}
C.~Angelantonj, Q.~Bonnefoy, C.~Condeescu and E.~Dudas,
``String Defects, Supersymmetry and the Swampland,''
[arXiv:2007.12722 [hep-th]].

\bibitem{Kuramochi2020}
R.~Kuramochi, S.~Mizoguchi and T.~Tani,
``Magic square and half-hypermultiplets in F-theory,''
[arXiv:2008.09272 [hep-th]].

\bibitem{He202009}
Y.~H.~He and A.~Lukas,
``Machine Learning Calabi-Yau Four-folds,''
[arXiv:2009.02544 [hep-th]].

\bibitem{Clingher2020}
A.~Clingher and A.~Malmendier,
``On K3 surfaces of Picard rank 14,''
[arXiv:2009.09635 [math.AG]].

\bibitem{AspinwallGross}P.~S.~Aspinwall and M.~Gross, ``The SO(32) heterotic string on a K3 surface'', {\it Phys. Lett.} {\bf B387} (1996) 735--742 [arXiv:hep-th/9605131].
\bibitem{AMrational}P.~S.~Aspinwall and D.~R.~Morrison, ``Nonsimply connected gauge groups and rational points on elliptic curves'', {\it JHEP} {\bf 9807} (1998) 012 [arXiv:hep-th/9805206].
\bibitem{MMTW}C.~Mayrhofer, D.~R.~Morrison, O.~Till and T.~Weigand, ``Mordell-Weil Torsion and the Global Structure of Gauge Groups in F-theory'', {\it JHEP} {\bf 10} (2014) 16 [arXiv:1405.3656 [hep-th]].

\bibitem{Cas}J.~W.~S.~Cassels, {\it Lectures on Elliptic Curves}, London Math. Society Student Texts {\bf 24}, Cambridge University Press (1991).

\bibitem{Kod1}K.~Kodaira, ``On compact analytic surfaces II'', {\it Ann. of Math.} {\bf 77} (1963), 563--626.
\bibitem{Kod2}K.~Kodaira, ``On compact analytic surfaces III'', {\it Ann. of Math.} {\bf 78} (1963), 1--40.
\bibitem{Ner}A.~N\'eron, ``Mod\`eles minimaux des vari\'et\'es ab\'eliennes sur les corps locaux et globaux'', {\it Publications math{\' e}matiques de l'IH{\' E}S} {\bf 21} (1964), 5--125.
\bibitem{Tate}J.~Tate, ``Algorithm for determining the type of a singular fiber in an elliptic pencil'', in {\it Modular Functions of One Variable IV}, Springer, Berlin (1975), 33--52.

\bibitem{Muk}S.~Mukai, {\it An introduction to invariants and moduli}, Cambridge University Press (2003).

\bibitem{Font2020}
A.~Font, B.~Fraiman, M.~Gra\~na, C.~A.~N\'u\~nez and H.~P.~De Freitas,
``Exploring the landscape of heterotic strings on $T^d$,''
{\it JHEP} \textbf{10}, 194 (2020)
[arXiv:2007.10358 [hep-th]].

\bibitem{Moore}G.~W.~Moore, ``Les Houches lectures on strings and arithmetic'', [arXiv:hep-th/0401049].

\bibitem{PS-S}I.~I.~Piatetski-Shapiro and I.~R.~Shafarevich, ``A Torelli theorem for algebraic surfaces of type K3'', {\it Izv. Akad. Nauk SSSR Ser. Mat.} {\bf 35} (1971), 530--572.
\bibitem{SI}T.~Shioda and H.~Inose, ``On Singular K3 surfaces'', in W.~L.~Jr.~Baily and T.~Shioda (eds.), {\it Complex analysis and algebraic geometry}, Iwanami Shoten, Tokyo (1977), 119--136.

\bibitem{ShimadaZhang}I.~Shimada and D.-Q.~Zhang, ``Classification of extremal elliptic K3 surfaces and fundamental groups of open K3 surfaces'', {\it Nagoya Math. J.} {\bf 161} (2001), 23--54, [arXiv:math/0007171].

\bibitem{Apruzzi201808}
F.~Apruzzi, F.~Hassler, J.~J.~Heckman and T.~B.~Rochais,
``Nilpotent Networks and 4D RG Flows,''
{\it JHEP} \textbf{05}, 074 (2019)
[arXiv:1808.10439 [hep-th]].

\bibitem{Carta201809}
F.~Carta, S.~Giacomelli and R.~Savelli,
``SUSY enhancement from T-branes,''
{\it JHEP} \textbf{12}, 127 (2018)
[arXiv:1809.04906 [hep-th]].

\end{thebibliography}
\end{document}